\documentclass[12pt,a4paper]{article}
\usepackage{amsmath}
\numberwithin{equation}{section}
\title{Application of the modified method of simplest equation for obtaining exact 
traveling-wave solutions for the extended Korteweg - de Vries equation and 
generalized Camassa-Holm equation}
\author{Nikolay K. Vitanov$^{1,2}$\footnote{e-mail: vitanov@imbm.bas.bg;
vitanov@mpipks-dresden.mpg.de},  Zlatinka I. Dimitrova$^{3}$, Holger Kantz $^{2}$}
\date{$^{1}$ Institute of Mechanics, Bulgarian Academy of Sciences,
Akad. G. Bonchev Str., Bl. 4, 1113 Sofia, Bulgaria \\
$^{2}$ Max-Planck-Institute for the Physics of Complex Systems,
Noethnitzerstr. 38, 01187 Dresden, Germany \\
$^{3}$ "G. Nadjakov" Institute for Solid State Physics, Bulgarian Academy 
of Science, Blvd. Tzarigradsko Chaussee 72, 1784 Sofia, Bulgaria}
\begin{document}
\maketitle
\begin{abstract}
The modified method of simplest equation is applied to the extended
Korteweg - de Vries equation and to generalized Camassa - Holm equation.
Exact traveling wave solutions of these two nonlinear partial differential equations are
obtained. The equations of Bernoulli, Riccati and the extended tanh - equation are
used as simplest equations. Some of the obtained solutions correspond to surface water 
waves. 
\end{abstract}
\begin{flushleft}
{\bf MSC 2010 numbers:}
\end{flushleft}
35Q35 : PDEs in connection with fluid mechanics \\
35Q53 : KdV-like equations \\
\begin{flushleft}
{\bf Key words:} 
\end{flushleft}
method of simplest equation, exact traveling-wave solutions,
generalized Camassa - Holm equation, extended Korteweg - de Vries equation,
water waves
\newpage
\section{Introduction}
Because of the nonlinearity of the Navier - Stokes equation, Boussinesq equations, etc.,
the hydrodynamics is a rich sorce of nonlinear problems \cite{c1} - \cite{c8}. In addition
in the last decades the nonlinear partial differential equations are widely used
for modeling natural and social phenomena and systems. Examples
are \cite{scot1, longtin} from neurology, \cite{ac} -\cite{newell}  from the theory of solitons,  
\cite{murr} from the biology, \cite{perko,may} from
theory of dynamical systems, chaos theory and ecology, and \cite{temam} - \cite{vx}
from hydrodynamics and theory of turbulence.   The knowledge of exact solutions of
the model PDEs is usefull in several directions. First of all such solutions describe different kinds 
of waves. In addition the exact solutions are useful  as initial condition for  computer programs that 
simulate processes in the investigated system. Because of this an important research area
is connected to obtaining exact analytical or approximate numerical
solutions of such model PDEs. Among the  most famous methods for obtaining
solutions of exact solvable equations are the inverse scattering transform
\cite{gardner} -  \cite{ablowitz2} and the method of Hirota \cite{hirota}.
In the course of years numerous approaches for obtaining exact special
solutions of nonlinear PDE have been developed \cite{kudr90} - \cite {vit11}.
By means of such methods numerous exact solutions of many equations
have been obtained  such as for an example the Kuramoto-Shivasinsky equation 
\cite{kudr05x}, equations, connected to the models of population dynamics 
\cite{zeppet} - \cite{vd10b}, sine - Gordon equation \cite{lou} - \cite{radha},
sinh-Gordon or Poisson - Boltzmann equation \cite{nakamura} - \cite{wazwaz06},
Lorenz - like systems \cite{vit07}, etc. 
\par
One of the dirct methods for obtaining exact and approximate solutions of nonlinear
partial differential equations is the method of simplest equation \cite{kudr05x} - \cite{vit11}.
This method ant its modification called modified method of simplest equation
have been applied to numerous nonlinear PDEs \cite{kudr08}, \cite{kudr07x}, such as
Fisher equation and Fisher - like equations \cite{vit10_1}, classes of reaction - diffusion and 
reaction - telegraph equations \cite{vd10a}, generalized Kuramoto - Sivashinsky equation 
\cite{vdk10}, generalized Swift - Hohenberg and generalized Rayleigh equations \cite{vit11},
generalized Deagsperis - Processi equation and $b$-equation \cite{vit11a}.
In this paper we shall apply the modified method of simplest equation to the
extended Korteweg - de Vries equation and to generalized Camassa - Holm equation.
The goal of the paper is to obtain exact traveling - wave solutions to the last two equations.
\par
The paper is organized as follows. In Sect. 2 information from the theory of water waves
is presented which connects the Navier - Stokes equation and the profile of a class of water
waves to the exact traveling - wave solutions of the extended Korteweg - de Vries equation and 
Camassa - Holm equation. Sect. 3 contains information about the modified method of simplest
equation  needed for obtaining exact traveling - wave solution of the extended Korteweg - de Vries
equation and of the generalized Camassa - Holm equation.  In sect 4. we obtain exact traveling-wave
solutions of the extended Korteweg - de Vries equation on the basis of the use of the equations
of Bernoulli, Riccati and extended tanh - equation as simplest equations. In section 5 using the
same procedure we obtain exact traveling - wave solutions of the generalized Camassa - Holm
equation. Several concluding remarks are summarized in Sect. 6.
\section{Water waves, extended Kortweg - de - Vries equation and generalized Camassa - Holm equation}
 The research on water waves is a rich sorce of nonlinear problems \cite{benney} - \cite{j1}. 
We shall obtain the extended Korteweg - de Vries equation and the Camassa - Holm equation
following Johnson  \cite{j2}. 
Let us consider incompressible inviscid fluid with zero surface tension. The fluid moves over
an steady intermediate bed with coordinate $z=z^*(x,y)$. The pressure on the top boundary 
of the fluid is constant: $P=P_a$, where $P_a$ is the atmospheric pressure. The system of 
model equations for description of the movement of the fluid consists of the Navier - Stokes
equation and continuity equation
\begin{equation}\label{ns}
\frac{\partial \vec{u}}{\partial t} + \vec{u} \cdot \nabla \vec{u} = - \frac{1}{\rho} \nabla P +
\vec{F}
\end{equation}
\begin{equation}\label{cont}
\nabla \cdot \vec{u} =0
\end{equation}
where $\vec{u} = (u,v,w)$ is the velocity of the fluid, $\rho$ is the density of the fluid
and the acting force is gravity: $\vec{F} = (0,0,-g)$. The fluid surface is at $z=h(x,y,t)$
and the boundary conditions are as follows. 
\begin{itemize}
\item
On the fluid surface
\begin{equation}\label{bc1}
P = P_a; \hskip.5cm w = \frac{\partial h}{\partial t} + (\vec{u} \cdot \nabla)h
\end{equation}
\item
On the bed $z=z^*(x,y)$
\begin{equation}
w = u \frac{\partial z^*}{\partial x} + v \frac{\partial z^*}{\partial y}
\end{equation}
\end{itemize}
Let $h_0$ be the undisturbed depth of the water. Then we introduce the non-dimensional
deviation $\eta$ from $h_0$ as follows
\begin{equation}\label{nond_height}
h = h_0 + a \eta
\end{equation}
In addition we introduce the non - dimensional pressure $p$
\begin{equation}\label{nond_press}
P = P_a + \rho g (h_0 - z) + \rho g h_0 p
\end{equation}
$p$ measures the deviation from the hydrostatic pressure. Our goal is to study wave
propagation on the fluid surface. The scale of this motion is the typical wavelegth
of the surface wave $\lambda$. We shall use $\lambda$ as unit for length.
\par
Next we perform a non-dimensionalization of the model equations by the following substitutions
($\to$ means "replace by" and in this way the notation is retained)
\begin{eqnarray}\label{nond1}
x \to \lambda x; \hskip.5cm y \to \lambda y; \hskip.5cm z \to h_0 z; \hskip.5cm t \to
\frac{\lambda}{\sqrt{g h_0} t}; \nonumber \\
u \to \sqrt{g h_0} u; \hskip.5cm v \to \sqrt{g h_0} v; \hskip.5cm w \to 
\frac{h_0 \sqrt{g h_0}}{\lambda} w; \hskip.5cm z^* \to h_0 z^*
\end{eqnarray}
In addition we introduce two important parameters: amplitude parameter $\epsilon$ and
shallowness parameter $\delta$:
\begin{equation}\label{def_par}
\epsilon = \frac{a}{h_0}; \hskip.5cm \delta = \frac{h_0}{\lambda}
\end{equation}
Further we rescale the non-dimensionalized equations with respect to the amplitude
parameter $\epsilon$:
\begin{equation}\label{eps_rescaling}
u \to \epsilon u; \hskip.5cm v \to \epsilon v; \hskip.5cm w \to \epsilon w; \hskip.5cm p \to 
\epsilon p
\end{equation}
and obtain the system of equations
\begin{equation}\label{mod1}
\frac{\partial u}{\partial t} + \epsilon (\vec{u} \cdot \nabla) u = - \frac{\partial p}{
\partial x}
\end{equation}
\begin{equation}\label{mod2}
\frac{\partial v}{\partial t} + \epsilon (\vec{u} \cdot \nabla) v = - \frac{\partial p}{
\partial y}
\end{equation}
\begin{equation}\label{mod3}
\nabla \cdot {\bf u} = 0
\end{equation}
with boundary conditions:
\begin{equation}\label{bc_1}
p = \eta; \hskip.5cm w = \frac{\partial \eta}{\partial t} + \epsilon \left( u 
\frac{\partial \eta}{\partial x} + v \frac{\partial \eta}{\partial y} \right) 
\end{equation}
on $z=1 + \epsilon \eta$ and
\begin{equation}\label{bc_2}
w = u \frac{\partial z^*}{\partial x} + v \frac{\partial z^*}{\partial y}
\end{equation}
on $z=z^*$.
We continue the scaling of  (\ref{mod1}) - (\ref{mod3})
as follows:  $\zeta = \sqrt{\epsilon} (x - t)$; $\tau = \epsilon \sqrt{\epsilon} t$;
$w = \sqrt{\epsilon} W$ and expand all quantities
\begin{equation}\label{decomp}
q \sim \sum_{n=0}^{\infty} \sum_{m=0}^{\infty} \epsilon^n \delta ^{2 m} q_{nm}
\end{equation}
Assuming that $\epsilon \to 0$ and $\delta \to 0$ (small amplitude water waves)
from (\ref{mod1}) - (\ref{mod3}) we obtain that the non-dimensionalized height of the
water surface $\eta = \eta_{00} + \epsilon_1 \eta_{10} + \delta^2 \eta_{01} + \epsilon
\delta^2 \eta_{11}$ satisifies the higher order Korteweg - de Vries equation
\begin{equation}\label{kdv}
2 \frac{\partial \eta}{\partial \tau} + 3 \eta \frac{\partial \eta}{\partial \zeta} + 
\frac{1}{3} \delta^2 \frac{\partial^3 \eta}{\partial \zeta^3}
-\frac{3}{4} \epsilon \eta^2 \frac{\partial \eta}{\partial \zeta} = - \frac{1}{12} \epsilon 
\delta^2
\left( 23 \frac{\partial \eta}{\partial \zeta} \frac{\partial^2 \eta}{\partial \zeta^2} + 10 \eta 
\frac{\partial^3 \eta}{\partial \zeta^3}
 \right) + O(\epsilon^2, \delta^4)
\end{equation}
and to the same order the solution is
\begin{equation}\label{sol_ch2}
u \approx \eta - \frac{1}{4} \eta^2 + \epsilon \delta^2  \left( \frac{1}{3} - \frac{1}{2} 
z^2\right) \frac{\partial^2 \eta}{\partial \zeta^2}
\end{equation}
where $z$ is connected to the depth of the moving fluid. Eq. (\ref{kdv}) is the extended
Korteweg - de Vries equation and it will be one of equations we shall discuss below.
\par
Another important water wave equation - the Camassa - Holm equation can be obtained in
similar way like (\ref{kdv}) above.
The Camassa - Holm equation describes the $x-$ component $u$ of the fluid
velocity at certain depth $z_0$ below the fluid surface.  Cammassa - Holm equation
was introduced as completely integrable bi-Hamiltonian dispersive shallow water equation \cite{ch1}.
 In the framework discussed here  and in \cite{j2} the Camassa - Holm equation is
\begin{equation}\label{ch}
\frac{\partial U}{\partial T} + 2 \kappa \frac{\partial U}{\partial X} + 
3 \epsilon U \frac{\partial U}{\partial X} - \epsilon \delta^2 
\frac{\partial^3 U}{ \partial X^2 \partial T}= \epsilon^2 \delta^2 
\left( 2 \frac{\partial U}{\partial X}\frac{\partial^2 U}{\partial X^2} + U 
\frac{\partial^3 U}{\partial X^3} \right )
\end{equation}
The error of this equation is $O(\epsilon^3, \epsilon \delta^4)$; it is valid for
$z_0 = \frac{1}{\sqrt{2}}$. In addition $\kappa = \frac{4}{5} \sqrt{\frac{3}{5}}$ 
and the solution of the Camassa - Holm equation equation corresponds within 
the approximation $O(\epsilon^3, \epsilon \delta^4)$ to the surface wave
\begin{equation}\label{ch_wave}
\eta \sim \sqrt{\frac{5}{3}} \left(U + \frac{1}{4} \sqrt{\frac{5}{3}} \epsilon U^2 
- \frac{1}{5} \epsilon \delta^2 \frac{\partial^2 U}{\partial X^2} \right) 
\end{equation}
where $T = t \sqrt{\epsilon}$; $X = 2 \sqrt{\frac{5}{3}} \left(Z - \frac{3}{5} T \right)$;
$Z=x \sqrt{\epsilon}$.
\par
In this paper we shall discuss the following generalized Camassa - Holm equation
\begin{eqnarray}\label{gen_ch}
\frac{\partial U}{\partial T} + p_1 \frac{\partial U}{\partial X} - 
p_2 \frac{\partial^3 U}{\partial X^2 \partial T} - p_3 \frac{\partial^3}{\partial X^3} f(U)
+ \nonumber \\
p_4 \frac{\partial}{\partial X} \left[ \frac{1}{2} g (U) + \frac{p_5}{2}
\frac{d^2 f}{d U^2} \left( \frac{\partial U}{\partial X}\right)^2 \right ] =0
\end{eqnarray}
where $p_1;p_2;p_3;p_4;p_5$ are parameters.
Let $f(U)= \frac{1}{2} U^2$; $g(U) = C + 3 U^2$ ($C$ is a constant) and $p_3 = p_4 p_5$. 
Then from Eq. (\ref{gen_ch}) we obtain
\begin{equation}\label{gen_ch1}
\frac{\partial U}{\partial T} + p_1 \frac{\partial U}{\partial X} - 
p_2 \frac{\partial^3 U}{\partial X^2 \partial T} + 3 p_4 U \frac{\partial U}{\partial X}
-2 p_3 \frac{\partial U}{\partial X} \frac{\partial^2 U}{\partial X^2}
- p_3 U \frac{\partial^3 U}{\partial X^3} =0
\end{equation}
Eq. (\ref{gen_ch}) contains the Camassa - Holm equation as particular case when
$p_2 = 2 \kappa$; $p_4 = \epsilon$; $p_2 = \epsilon \delta^2$; $p_3 =
\epsilon^2 \delta^2$. In general in Eq. (\ref{gen_ch}) $f(U)$ and $g(U)$ can
be arbitrary polynomials of $U$. Finally we note that if we do not set $g(U)$ and
all other settings are as for obtaining Eq. (\ref{gen_ch1}) from Eq. (\ref{gen_ch})
we obtain the following generalization of the Camassa - Holm equation
\begin{equation}\label{gen_ch2}
\frac{\partial U}{\partial T} + p_1 \frac{\partial U}{\partial X} +
\frac{p_4}{2} \frac{\partial}{\partial X} g(U) - p_2 \frac{\partial ^3 U}{\partial X^2 
\partial T} - 2 p_3 \frac{\partial U}{\partial X} \frac{\partial^2 U}{\partial X^2}
- p_3 U \frac{\partial^3 U}{\partial X^3} =0
\end{equation}
Let $p_2 = p_4 =1$;$p_1 = p_3$. Then we obtain the version of the generalized
Camassa - Holm equation discussed by Tian and Wang \cite{tw}  ( see also \cite{ts, ty}).
\section{The modified method of simplest equation}
The method of simplest equation has been developed by Kudryashov \cite{kudr05x, kudr05}
on the basis of a procedure analogous to the first step
of the test for the Painleve property. In the modified method of the simplest
equation \cite{vdk10, vit11} this procedure is substituted by  the concept for
the balance equation (for details about the balance equation see the  subsection of this section).
The essence of the method is as follows.
Let us have a partial differential equation and let by means of an
appropriate ansatz this equation is reduced to a nonlinear
ordinary differential equation
\begin{equation}\label{gen_eq}
P \left( u(\xi),\frac{d u}{d \xi},\frac{d^{2} u}{d \xi^{2}},\dots \right) = 0
\end{equation}
For large class of equations from the kind (\ref{gen_eq}) exact solution 
can be constructed as finite series
\begin{equation}\label{solution}
u(\xi) = \sum_{\mu=-\nu}^{\nu_1} \theta_{\mu} [V (\xi)]^{\mu}
\end{equation}
where  $\nu>0$, $\mu >0$, $\theta_{\mu}$ is a parameter and $V(\xi)$ is a solution of 
some ordinary differential equation
referred to as the simplest equation. The simplest equation is of lesser 
order than (\ref{gen_eq}) and we know the general solution of
the simplest equation or we know at least exact analytical particular
solution(s) of the simplest equation \cite{kudr08, kudr07x}. 
\par
The  modified method of simplest equations can be applied to
nonlinear partial differential equations of the kind
\begin{equation}\label{basic_eq1}
F \left(\frac{\partial ^{\omega_1} u}{\partial x ^{\omega_1}},
\frac{\partial ^{\omega_2} u}{\partial t ^{\omega_2}}, 
\frac{\partial ^{\omega_3} u}{\partial x ^{\omega_4} \partial t^{\omega_5}}\right) = G(u)
\end{equation}
where $\omega_3 = \omega_4 + \omega_5$ and
\begin{enumerate}
\item
$\frac{\partial ^{\omega_1} u}{\partial x ^{\omega_1}}$ denotes the set of derivatives
$$
\frac{\partial ^{\omega_1} u}{\partial x ^{\omega_1}} = \left( \frac{\partial u}{\partial 
x}, \frac{\partial^2 u}{\partial x^2}, \frac{\partial u^3}{\partial x^3}, \dots \right)
$$
\item
$\frac{\partial ^{\omega_2} u}{\partial t ^{\omega_2}}$ denotes the set of derivatives
$$
\frac{\partial ^{\omega_2} u}{\partial t ^{\omega_2}} = \left( \frac{\partial u}{\partial 
t}, \frac{\partial^2 u}{\partial t^2}, \frac{\partial u^3}{\partial t^3}, \dots \right)
$$
\item
$\frac{\partial^{\omega_3} u}{\partial x^{\omega_4} \partial t^{\omega_5}}$ denotes the set
of derivatives
$$
\frac{\partial^{\omega_3} u}{\partial x^{\omega_4} \partial t^{\omega_5}} = \left( 
\frac{\partial^2 u}{\partial x \partial t}, \frac{\partial^3 u}{\partial x^2 \partial t},
\frac{\partial u^3}{\partial x \partial t^2}, \dots  \right)
$$
\item
$G(u)$ can be
\begin{enumerate}
\item
polynomial of $u$ or
\item
function of $u$ which can be reduced to polynomial of $u$ by means of
Taylor series for small values of $u$
\end{enumerate}
\item
$F$ can be an arbitrary sum of products of arbitrary number of its arguments.
Each argument in each product can have arbitrary power. Each of the products
can be multiplied by a function of $u$ which can be
\begin{enumerate}
\item
polynomial of $u$ or
\item
function of $u$ which can be reduced to polynomial of $u$ by means of
Taylor series for small values of $u$
\end{enumerate}
\end{enumerate}
For this class of equations the modified method of simplest equation allows us in principle 
to search for
\begin{enumerate}
\item
Exact traveling-wave solutions of (\ref{basic_eq1}) if $G(u)$ and the multiplication functions
form item 5. above are polynomials 
\item
Approximate traveling-wave solutions for small $u$ in all other cases
\end{enumerate}
\par
We are interested in traveling wave solutions
\begin{equation}\label{tr_wav}
u(x,t)=u(\xi) = u(x - vt)
\end{equation}
of the extended Korteweg - de vries equation and of the generalized Camassa - Holm equation. 
These solutions will be obtained on the basis of the modified method of simplest equation.
\par
In what follows below we shall set $\nu =0$ and we shall use as simplest equation 
particular cases of the following equation
\begin{equation}\label{simplest}
\frac{d V}{d \xi} = \sum_{\alpha=0}^{\beta} \gamma_{\alpha} [V(\xi)]^{\alpha}
\end{equation} 
where $\gamma_{\alpha}$ is a parameter. Two particular cases of (\ref{simplest})
are the equations of Bernoulli and Riccati 
which are well known nonlinear ordinary differential
equations and their solutions can be expressed by elementary functions.
For the Bernoulli equation $\gamma_0 = 0$; $\gamma_1 = a$; $\gamma_2 = \dots
= \gamma_{k-1} =0$; $\gamma_k =c$; $\gamma_{k+1}= \dots = \gamma_{\beta} =0$.
The result is
\begin{equation}\label{bernoulli}
\frac{d V}{d \xi} = a V(\xi) + c [V (\xi)]^{k}
\end{equation}
where $k$ is an integer and $k>1$. For (\ref{bernoulli}) we shall use the solutions
\begin{equation}\label{bsol1}
V (\xi) = \left \{\frac{a \exp [a(k-1)(\xi+ \xi_{0})]}{1-
c \exp[a(k-1)(\xi+\xi_{0})]} \right \}^{\frac{1}{k-1}}
\end{equation}
for the case $c<0$, $a>0$ and
\begin{equation}\label{bsol2}
V(\xi) = \left \{ -\frac{a \exp [a(k-1)(\xi+\xi_{0})]}{1+
c \exp[a(k-1)(\xi+\xi_{0})]} \right \}^{\frac{1}{k-1}}
\end{equation}
for the case $c>0$, $a<0$. Above $\xi_{0}$ is a constant of
integration. For the Riccati equation $\gamma_0 = d$; $\gamma_1 = c$; $\gamma_2 = a$,
$\gamma_3 = \dots=\gamma_{\beta} =0$.
The result is
\begin{equation}\label{riccati}
\frac{d V}{d \xi} = a [V (\xi)]^{2} + c V (\xi) + d
\end{equation}
We shall use two solutions of (\ref{riccati}). By a substitution in (\ref{riccati})
it can be shown that
\begin{equation}\label{sol_ric1}
V_1(\xi) = - \frac{c}{2a} - \frac{\theta}{2a} \tanh
\left[\frac{\theta (\xi + \xi_0)}{2} \right]
\end{equation}
is a particular solution of the Riccati equation. On the basis of (\ref{sol_ric1})
we can construct more complicated solution of the Riccati equation which has the
form
$$
V_2 (\xi) = V_1 (\xi) + \frac{1}{V_3 (\xi)}
$$
where $V_3(\xi)$ is solution of the linear equation
$$
\frac{d V_3}{d \xi} + [c + 2 a V_1(\xi)] V_3(\xi) = -a
$$
For the solution $V_2(\xi)$ we obtain 
\begin{eqnarray}\label{sol_ric2}
V_2(\xi) = - \frac{c}{2a} - \frac{\theta}{2a} \tanh
\left(\frac{\theta (\xi + \xi_0)}{2} \right) + \nonumber \\
\cfrac{\exp \left(\cfrac{\theta (\xi+\xi_0)}{2}
\right)}{ 2 \cosh \left(\cfrac{\theta (\xi +\xi_0)}{2} \right) \left[\cfrac{a}{\theta}+
2 C^{*} \exp \left(\cfrac{\theta (\xi + \xi_0)}{2} \right) \cosh \left(\cfrac{\theta
(\xi + \xi_0)}{2} \right)  \right]} \nonumber \\
\end{eqnarray}
In (\ref{sol_ric1}) and (\ref{sol_ric2}) $\theta^{2} = c^{2} - 4ad >0$.
In addition $\xi_0$ and $C^{*}$ are  constants of integration. We shall
use the solutions $V_1(\xi)$ and $V_2(\xi)$ below.
\par
As particular case of the use of equation of Riccati as simplest equation
we shall consider also the so called  extended tanh-function equation 
\begin{equation}\label{eth}
\frac{d V}{d \xi} = \overline{d}^{2} - 
\overline{a}^{2} V^{2}  
\end{equation}
used by Fan \cite{fan00} as basis for the extended tanh-function method.
(\ref{eth}) is obtained from (\ref{riccati}) when $b=0$, 
$a = -\overline{a}^{2}$, and $d = \overline{d}^{2}$. The solution of
(\ref{eth}) we shall use below is
\begin{equation}\label{sol_eth}
V (\xi) = \frac{\overline{d}}{\overline{a}} \tanh [\overline{a} \ \overline{d}
(\xi + \xi_{0})]
\end{equation}
where $\overline{a}^2 V (\xi)^{2} < \overline{d}^{2}$ and $\xi_{0}$ is a constant
of integration. The maximum power of $V(\xi)$ in (\ref{eth}) is the same
as the maximum power of $V(\xi)$ in (\ref{riccati}). Because of this the
obtained below balance equations for the equation of Riccati and for 
the extended tanh-equation will be the same. The difference between
the two simplest equations is that the equation of Riccati will lead to
more complicated exact solutions of the corresponding PDE. 
\subsection{Balance equations} 
Let us search for a solution of (\ref{gen_eq}) which is of the 
kind (\ref{solution}) with $\eta > \nu$ and where $V (\xi)$ is a solution of (\ref{simplest}). 
The substitution of (\ref{solution}) and 
the corresponding simplest equation
in (\ref{gen_eq}) leads to a polynomial equation of the kind
\begin{equation}\label{pol_eq}
P = \kappa_{r} V^{r} + \kappa_{r-1} V^{r-1} + \dots + \kappa_{0}=0
\end{equation}
where $r$ is some integer and the coefficients $\kappa$ depend on the
parameters of the solved equation as well as on the parameters of the solution.
The solution of (\ref{pol_eq}) is obtained when all coefficients in (\ref{pol_eq})
are equal to 0. This leads  to a system of nonlinear algebraic relationships
\begin{equation}\label{nonl_syst}
\kappa_{l}=0,\hskip.25cm  l=r,r-1,\dots,0
\end{equation}
In the classic version of the simplest equation method Kudryashov
determines $\nu_1$ by a procedure analogous to the first step of the
test for the Painleve property \cite{kudr05}. Below
we shall follow another idea \cite{vdk10, vit11}. We have to ensure that there are at least
two terms containing $V^r$ in (\ref{pol_eq}). In order to achieve this
we have to balance at least two of the  largest powers
of the polynomials arising from the terms of  (\ref{gen_eq}). Below we shall do this
for the extended Korteweg - de Vries equation and for the generakized Camassa - Holm equation.
\par
Let us now conseder the extended Kortweg - de vries equation (\ref{kdv})
As first we shall use the equation of Bernoulli (\ref{bernoulli}) as simplest equation.
The solution of Eq. (\ref{kdv}) will be constructed by means of solutions of Eq. 
(\ref{bernoulli}) on the basis of (\ref{solution}) with $\nu =0$. We shall solve generalized 
version of the extended Korteweg - de Vries equation
\begin{equation}\label{gen_kdv}
p_1 \frac{\partial \eta}{\partial \tau} + p_2 \eta \frac{\partial \eta}{\partial \zeta} + 
p_3 \frac{\partial^3 \eta}{\partial \zeta^3}
+ p_4 \eta^2 \frac{\partial \eta}{\partial \zeta} + p_5  \frac{\partial \eta}{\partial \zeta} 
\frac{\partial^2 \eta}{\partial \zeta^2} + p_6 \eta  \frac{\partial^3 \eta}{\partial \zeta^3}
=0
\end{equation}
Eq (\ref{gen_kdv}) is reduced to Eq. (\ref{kdv}) for the following values of the parameters
\begin{equation}\label{parameters}
p_1 = 2; \hskip.25cm p_2 =3; \hskip.25cm p_3 = \frac{1}{3} \delta^2; \hskip.25cm
p_4 = - \frac{3}{4} \epsilon; \hskip.25cm p_5 = \frac{23}{12} \epsilon \delta^2;
\hskip.25cm p_6 = \frac{5}{6} \epsilon \delta^2
\end{equation}
We shall introduce the coordinate $\xi = \zeta - v \tau$ and shall search for
traveling-wave solutions of Eq. (\ref{gen_kdv}). In the original coordinates $x$
and $t$ : $\xi = \sqrt{\epsilon} (x - \epsilon v t)$. 
The application of the modified method of simplest equation to Eq. (\ref{gen_kdv})
on the basis of equation of Bernoulli as simplest equation leads to the following
balance equation
\begin{equation}\label{kdv_bern_balance}
\nu_1 = 2 (k-1)
\end{equation}
\par
Let now the equation of Riccati (\ref{riccati}) be used as simplest equation.
We again search solution of (\ref{gen_kdv}) in the form (\ref{solution}) with $\nu =0$.
For this case the balance equation is
\begin{equation}\label{kdv_riccati_balance}
\nu_1 = 2
\end{equation}
The use of the extended tanh - equation as simplest equation leads to the same balance equation as for the
case of Riccati equation namely (\ref{kdv_riccati_balance}).
\par
Now let us discuss the generalized Camassa - Holm equation (\ref{gen_ch}).
We are interested in traveling wave solutions. Because of this we introduce the
coordinate $\xi = X - v T$ where $v$ is the velocity of the corresponding wave.
First let the equation of Bernoulli (\ref{bernoulli}) be used as simplest equation
and the solution of Eq. (\ref{gen_ch}) be searched in the form of Eq. (\ref{solution}) with $\nu =0$. 
The determination of the maximum powers of the terms in Eq. (\ref{gen_ch}) leads to
the following balance equation 
\begin{equation}\label{bal_ch1}
\nu_1 = \frac{2 (k-1)}{J - I}
\end{equation}
where $k>1$ and $J>I>1$. For $I>1$ another balance equation is possible namely 
\begin{equation}\label{bal_ch1a}
I \nu_1 +3 (k-1) = I \nu_1 + 3 (k-1)
\end{equation}
when $(J-I) \nu_1 < 2 (k-1)$. This balance is possible as there are two terms in
(\ref{gen_ch}) which lead to maximum power $I \nu_1 + 3(k -1)$ of terms of the
resulting polynomial and these powers are maximum powers for the entire polynomial
when $(J-I) \nu_1 < 2 (k-1)$.
\par
Finally if $I=1$  the balance equation is
\begin{equation}\label{bal_ch1b}
\nu_1 = \frac{2 (k-1)}{J-1}
\end{equation} 
\par
Another possibility is to use the equation of Riccati (\ref{riccati}) as simplest
equation. The solution of Eq. (\ref{gen_ch}) is searched again the form of 
Eq. (\ref{solution}). Several balance equations are possible for this case.
They are
\begin{equation}\label{bal_ch2}
\nu_1 = I ; \hskip.5cm J < \nu_1 + 2; \hskip.5cm J < I +2
\end{equation}
\begin{equation}\label{bal_ch3}
\nu_1 +2  = J ; \hskip.5cm I < \nu_1 ; \hskip.5cm I+2 < J
\end{equation}
\begin{equation}\label{bal_ch4}
J  = I + 2 ; \hskip.5cm \nu_1 < I ; \hskip.5cm \nu_1 + 2  < J
\end{equation}
The possible balance equations for the case when the extended tanh - equation is used as
simplest equation are the same as for the case of Riccati equation.
\par
Finally we would like to note two of the papers
of Kudryashov \cite{k1, k2} who advises us to be careful when 
use the methods for obtaining of exact solutions of the nonlinear differential equations.
Especially we have to check if the newly obtained solutions can be reduced to
some of the already known exact solutions of the correspondent PDE.
\section{Exact traveling wave solutions of the extended Korteweg - de Vries equation (\ref
{kdv})}
First of all we shall use the equation of Bernoulli as simplest equation.
Let $k=2$. Then from Eq. (\ref{kdv_bern_balance}) $\nu_1 =2$. The subsitution of Eq. 
(\ref{solution}) and Eq. (\ref{bernoulli}) in Eq. (\ref{gen_kdv}) leads to the following
system of $7$ nonlinear algebraic relationships between the parameters of Eq. (\ref{gen_kdv})
and parameters of its solution
\begin{eqnarray}\label{sys1}
0 &=& c \theta_2^2 [6 p_5 c^2 +  p_4 \theta_2 + 12 p_6  c^2]  \nonumber \\
0& =& \theta_2 [24 p_6 \theta_1  c^3 + p_6  (6 \theta_1 c^3 + 54\theta_2 a c^2) + 6 p_5 (
\theta_1 c + 2 \theta_2 a)  c^2 + \nonumber \\ && 2 p_5  c (10 \theta_2 a c + 2 \theta_1 c^2) 
+ 4 p_4 \theta_1 \theta_2 c + p_4 \theta_2 (\theta_1 c + 2 \theta_2 a)] 
\nonumber \\
0 &=& 24 p_3 \theta_2 c^3 + 6 p_5 \theta_1 a \theta_2 c^2 + p_5 (\theta_1 c + 2 \theta_2 a) 
(10 \theta_2 a c + 2 \theta_1 c^2) +  \nonumber \\ &&
2 p_5 \theta_2 c (3 \theta_1 a c + 4 \theta_2 a^2) + 2 p_2 
\theta_2^2 c + 2 p_4 (2 \theta_0 \theta_2 + \theta_1^2) \theta_2 c + \nonumber \\ &&
2 p_4 \theta_1 \theta_2 (\theta_1 c + 2 \theta_2 a) + 
p_4 \theta_2^2 \theta_1 a + 24 p_6 \theta_0 \theta_2 c^3 + 
 p_6 \theta_1 (6 \theta_1 c^3 + 54 \theta_2 a c^2) +  \nonumber \\ &&
p_6 \theta_2 (12 \theta_1 c^2 a + 38 \theta_2 a^2 c) \nonumber \\
0 &=& 4 p_4 \theta_0 \theta_1 \theta_2 c + p_4 (2 \theta_0 \theta_2 + \theta_1^2) (\theta_1 c 
+ 2 \theta_2 a) + 2 p_4 \theta_1^2 \theta_2 a + \nonumber \\ &&
 2 p_2 \theta_1 \theta_2 c + p_2 \theta_2 (\theta_1 c + 2 \theta_2 a) + p_3 (6 \theta_1 c^3 + 
54 \theta_2 a c^2) +  \nonumber \\ &&
p_6 \theta_0 (6 \theta_1 c^3 + 54 \theta_2 a c^2) + 
 p_6 \theta_1 (12 \theta_1 c^2 a + 38 \theta_2 a^2 c) +  \nonumber \\ &&
 p_6 \theta_2 (7 \theta_1 a^2 c + 8 
\theta_2 a^3) + p_5 \theta_1 a (10 \theta_2 a c + 2 \theta_1 c^2) +  \nonumber \\ &&
p_5 (\theta_1 c + 2 \theta_2 a) (3 \theta_1 a c + 4 \theta_2 a^2) + 2 p_5 \theta_2 c \theta_1 
a^2  \nonumber \\
0&=& - 2 v p_1 \theta_2 c + p_6 \theta_0 (12 \theta_1 c^2 a + 38 \theta_2 a^2 c) + 
p_6 \theta_1 
(7 \theta_1 a^2 c + 8 \theta_2 a^3) + \nonumber \\ &&
p_6 \theta_2 \theta_1 a^3 + 2 p_2 \theta_0 \theta_2 c + p_2 \theta_1 (\theta_1 c + 2 \theta_2 
a) + p_2 \theta_2 \theta_1 a +  \nonumber \\ &&
p_3 (12 \theta_1 c^2 a + 38 \theta_2 a^2 c) + 2 p_4 \theta_0^2 \theta_2 c + 2 p_4 \theta_0 
\theta_1 (\theta_1 c + 2 \theta_2 a) + \nonumber \\ &&
 p_4 (2 \theta_0 \theta_2 \theta_1^2) \theta_1 a + p_5 \theta_1 a (3 \theta_1 a c + 4 \theta_2 
a^2) + p_5 (\theta_1 c + 2\theta_2 a) \theta_1 a^2 \nonumber \\
0&=& - v p_1 (\theta_1 c + 2 \theta_2 a) + p_4 \theta_0^2 (\theta_1 c + 2 \theta_2 a) + 2 p_4 
\theta_0 \theta_1^2 a + p_5 \theta_1^2 a^3 + \nonumber \\&&
 p_2 \theta_0 (\theta_1 c + 2 \theta_2 a) + p_2 \theta_1^2 a + p_3 (7 \theta_1 a^2 c + 8 
\theta_2 a^3) + p_6 \theta_0 (7 \theta_1 a^2 c + 8 \theta_2 a^3) + \nonumber \\&&
 p_6 \theta_1^2 a^3 \nonumber \\
0 &=& \theta_1 [ - v p_1  a + p_4 \theta_0^2 a + p_6 \theta_0  a^3 + p_2 \theta_0 
 a + p_3  a^3] \nonumber \\
\end{eqnarray}
One solution of this system of algebraic equations is
\begin{eqnarray}\label{sol_sys1}
\theta_2 &=& - \frac{6 c^2 (p_5 + 2 p_6)}{p_4} ; \hskip.5cm \theta_1 = - \frac{6 a c (p_5 + 2 
p_6)}{p_4} \nonumber \\
\theta_0 &=& - \frac{-2 p_3 p_4 + a^2 p_5^2 + 3 a^2 p_5 p_6 + p_2 p_5 + 2 p_2 p_6 + 2 a^2 
p_6^2}{2 p_4 (p_5 + p_6)} \nonumber \\
v &=& \frac{a^4 p_5^4 + 4 p_3^2 p_4^2 - 4 p_2 p_3 p_4 p_6 - p_2^2 p_5^2 - 2 p_2^2 p_5 p_6 + 4 
a^4 p_6 p_5^3 + 5 a^4 p_6^2 p_5^2 + 2 a^4 p_6^3 p_5}{4 p_1 p_4 (p_5 + p_6)^2} \nonumber \\
\end{eqnarray}
The solution of Eq. (\ref{gen_kdv}) is
\begin{eqnarray}\label{gen_kdv_sol1}
\eta = - \frac{-2 p_3 p_4 + a^2 p_5^2 + 3 a^2 p_5 p_6 + p_2 p_5 + 2 p_2 p_6 + 2 a^2 
p_6^2}{2 p_4 (p_5 + p_6)} - \nonumber \\
 \frac{6 a^2 c (p_5 + 2 p_6)}{p_4} \left [ \frac{\exp [a(\xi+ \xi_{0})]}{1-
c \exp[a(\xi+\xi_{0})]} \right ] \left \{ 1 + c 
\left [ \frac{ \exp [a(\xi+ \xi_{0})]}{1-
c \exp[a(\xi+\xi_{0})]} \right ] \right \} \nonumber \\
\end{eqnarray}
for the case $a>0$, $c<0$ and 
\begin{eqnarray}\label{gen_kdv_sol2}
\eta = - \frac{-2 p_3 p_4 + a^2 p_5^2 + 3 a^2 p_5 p_6 + p_2 p_5 + 2 p_2 p_6 + 2 a^2 
p_6^2}{2 p_4 (p_5 + p_6)} + \nonumber \\
 \frac{6 a^2 c (p_5 + 2 p_6)}{p_4} \left [ \frac{ \exp [a(\xi+ \xi_{0})]}{1+
c \exp[a(\xi+\xi_{0})]} \right ] \left \{ 1   -  
c \left [ \frac{ \exp [a(\xi+ \xi_{0})]}{1+
c \exp[a(\xi+\xi_{0})]} \right ]\right \} \nonumber \\
\end{eqnarray}
for the case $a<0$, $c>0$. Let us discuss the case $a>0$; $c<0$. 
From Eq. (\ref{parameters}) we obtain the following solution of the extended Korteweg - de 
Vries equation (\ref{kdv})  
\begin{eqnarray}\label{kdv_wave}
\eta = \frac{540 + 473 \epsilon \delta^2 a^2}{198 \epsilon} + \nonumber \\
 \frac{86 \delta^2 c a^2}{3} \left [ \frac{\exp [a(\xi+ \xi_{0})]}{1-
c \exp[a(\xi+\xi_{0})]} \right ] \left \{ 1 + c 
\left [ \frac{ \exp [a(\xi+ \xi_{0})]}{1-
c \exp[a(\xi+\xi_{0})]} \right ] \right \} \nonumber \\
\end{eqnarray}
For the surface wave from Eq. (\ref{sol_ch2}) we obtain
\begin{eqnarray}\label{wave1}
u \approx \frac{540 + 473 \epsilon \delta^2 a^2}{198 \epsilon} + \nonumber \\
 \frac{86 \delta^2 c a^2}{3} \left [ \frac{\exp [a(\xi+ \xi_{0})]}{1-
c \exp[a(\xi+\xi_{0})]} \right ] \left \{ 1 + c 
\left [ \frac{ \exp [a(\xi+ \xi_{0})]}{1-
c \exp[a(\xi+\xi_{0})]} \right ] \right \} \nonumber \\
-\frac{1}{4} \bigg \{ \frac{540 + 473 \epsilon \delta^2 a^2}{198 \epsilon} + \nonumber \\
 \frac{86 \delta^2 c a^2}{3} \left [ \frac{\exp [a(\xi+ \xi_{0})]}{1-
c \exp[a(\xi+\xi_{0})]} \right ] \left \{ 1 + c 
\left [ \frac{ \exp [a(\xi+ \xi_{0})]}{1-
c \exp[a(\xi+\xi_{0})]} \right ] \right \}  \bigg\}^2 \nonumber \\
+ \epsilon \delta^2 \left(\frac{1}{3} - \frac{1}{2} z^2 \right) u^*(\xi) 
 \nonumber \\
\end{eqnarray}
where $u^*(\xi)$ is
\begin{eqnarray}\label{u_star}
u^* = \frac{86}{3} \delta^2 c a^4 \bigg [  \frac{ \exp [a (\xi + \xi_0)]}{1 - c \exp [a (\xi + 
\xi_0)]} + 21 c \frac{ \exp [a (\xi + \xi_0)]^2}{\{1 - c \exp [a (\xi +
\xi_0)]\}^2} + \nonumber \\
12 c^2 \frac{ \exp [a (\xi + \xi_0)]^3}{ \{ 1 - c \exp [a (\xi +
\xi_0)] \}^3} + 
6 c^3 \frac{ \exp [a (\xi + \xi_0)]^3}{\{1 - c \exp [a (\xi +
\xi_0)]\}^4} \bigg ]
\end{eqnarray}
\par
Let now the equation of Riccati (\ref{riccati}) be used as simplest equation.
We again search solution of (\ref{gen_kdv}) in the form (\ref{solution}) with $\nu =0$.
For this case the balance equation is (\ref{kdv_riccati_balance})
The application of the modified method of simplest equation leads to a system of
8 algebraic relationships among the parameters of the extended Korteweg - de Vries equation
and the parameters of the solution. One solution of this nonlinear system of
algebraic relationships is
\begin{eqnarray}\label{sol_sys2}
\theta_2 &=& - \frac{6 a^2 (p_5 + 2 p_6)}{p_4} ; \hskip.5cm \theta_1 = - \frac{6 a c (p_5 + 2 
p_6)}{p_4} \nonumber \\
\theta_0 &=& - \frac{3 p_5 p_6 c^2 + 16 p_6^2 a d + 8 p_5^2 a d + 24 p_5 p_6 a d + p_2 p_5 + 2 
p_2 p_6 + p_5^2 c^2 + 2 p_6^2 c^2 - 2 p_3 p_4}{2 p_4 (p_5 + p_6)} \nonumber \\
v &=& \frac{v_1}{v_2} \nonumber \\
v_1&=& \frac{1}{4} ( - 4 p_2 p_6 p_3 p_4 - 40 p_5^2 a c^2 d p_6^2 - 16 p_5 a c^2 d p_6^3 + 4 
p_5^3 c^4 p_6 + 5 p_5^2 c^4 p_6^2 + 2 p_5 p_6^3 c^4 + \nonumber \\
&& 16 p_5^4 a^2 d^2 - 32 p_5^3 a c^2 d p_6 - 8 p_5^4 a c^2 d + 64 p_5^3 a^2 d^2 p_6 + 80 
p_5^2 a^2 d^2 p_6^2 + 32 p_5 p_6^3 a^2 d^2 + \nonumber \\
&& p_5^4 c^4 - 2 p_6 p_2^2 p_5 - p_2^2 p_5^2 + 4 p_3^2 p_4^2) 
\nonumber \\
v_2 &=& p_1 p_4(p_5+p_6)^2 \nonumber \\
\end{eqnarray}
The solution of Eq. (\ref{gen_kdv})  based on the solution (\ref{sol_ric1}) of the
equation of Riccati is 
\begin{eqnarray}\label{gen_kdv_sol3}
\eta = - \frac{3 p_5 p_6 c^2 + 16 p_6^2 a d + 8 p_5^2 a d + 24 p_5 p_6 a d + p_2 p_5 + 2 
p_2 p_6 + p_5^2 c^2 + 2 p_6^2 c^2 - 2 p_3 p_4}{2 p_4 (p_5 + p_6)} - \nonumber \\
\frac{6 a c (p_5 + 2 p_6)}{p_4} \left \{ - \frac{c}{2a} - \frac{\theta}{2a} \tanh
\left[\frac{\theta (\xi + \xi_0)}{2} \right] \right \} \left \{ 1+ \frac{a}{c}  
\left \{ - \frac{c}{2a} - \frac{\theta}{2a} \tanh
\left[\frac{\theta (\xi + \xi_0)}{2} \right] \right \}\right \} \nonumber \\
\end{eqnarray}
The solution of Eq. (\ref{gen_kdv})  based on the solution (\ref{sol_ric2}) of the
equation of Riccati is
\begin{eqnarray}\label{gen_kdv_sol4}
\eta = - \frac{3 p_5 p_6 c^2 + 16 p_6^2 a d + 8 p_5^2 a d + 24 p_5 p_6 a d + p_2 p_5 + 2 
p_2 p_6 + p_5^2 c^2 + 2 p_6^2 c^2 - 2 p_3 p_4}{2 p_4 (p_5 + p_6)} - \nonumber \\
 \frac{6 a c (p_5 + 2 p_6)}{p_4}  
\bigg \{ - \frac{c}{2a} - \frac{\theta}{2a} \tanh
\left(\frac{\theta (\xi + \xi_0)}{2} \right) + \nonumber \\
\cfrac{\exp \left(\cfrac{\theta (\xi+\xi_0)}{2}
\right)}{ 2 \cosh \left(\cfrac{\theta (\xi +\xi_0)}{2} \right) \left[\cfrac{a}{\theta}+
2 C^{*} \exp \left(\cfrac{\theta (\xi + \xi_0)}{2} \right) \cosh \left(\cfrac{\theta
(\xi + \xi_0)}{2} \right)  \right]} \bigg \} \times \nonumber \\
\bigg \{ 1 + \frac{a}{c} \bigg \{ - \frac{c}{2a} - \frac{\theta}{2a} \tanh
\left(\frac{\theta (\xi + \xi_0)}{2} \right) + \nonumber \\
\cfrac{\exp \left(\cfrac{\theta (\xi+\xi_0)}{2}
\right)}{ 2 \cosh \left(\cfrac{\theta (\xi +\xi_0)}{2} \right) \left[\cfrac{a}{\theta}+
2 C^{*} \exp \left(\cfrac{\theta (\xi + \xi_0)}{2} \right) \cosh \left(\cfrac{\theta
(\xi + \xi_0)}{2} \right)  \right]} \bigg \} \bigg \} \bigg \} \nonumber\\
\end{eqnarray}
We remember that in Eq. (\ref{gen_kdv_sol3}) and Eq. (\ref{gen_kdv_sol4}) 
$\theta^2 = c^2 - 4 a d >0$.
\par
Taking into account the relationships (\ref{parameters}) for
the surface wave connected to the solution (\ref{gen_kdv_sol3}) is 
\begin{eqnarray}\label{wave2}
u &\approx& -\frac{16}{1089 \epsilon^4 \delta^4} \bigg \{ 
\bigg (-\frac{15609}{64} \epsilon^5 \delta^7 \theta^4 + \frac{223729}{256} \epsilon^4 \delta^8 
\theta^4 + \frac{46827}{128} \epsilon^5 \delta^7 \theta^4 z^2 \bigg) \times \nonumber \\ 
&& \bigg [\tanh \bigg(\frac{1}{2} \theta (\xi+\xi_0)\bigg ) \bigg]^4 - 
\bigg ( \frac{21285}{32} \epsilon^3 \delta^6 \theta^2 - \frac{15609}{32} \epsilon^5 
\delta^7 
\theta^4 z^2 + \frac{223729}{48} \epsilon^4 \delta^8 a d \theta^2 - \nonumber \\
&& \frac{223729}{192} 
\epsilon^4 \delta^8 c^2 \theta^2 + \frac{5203}{16} \epsilon^5 \delta^7 \theta^4 - 
 \frac{15609}{32} \epsilon^4 \delta^6 \theta^2 \bigg) \bigg [\tanh \bigg( \frac{1}{2} \theta 
(\xi+\xi_0) \bigg) \bigg ]^2 -\nonumber \\
&& \bigg (
- \frac{1485}{8} \epsilon^3 \delta^4 + \frac{2025}{16} \epsilon^2 \delta^4 + 
\frac{223729}{576} \epsilon^4 \delta^8 c^4 + \frac{5203}{16} \epsilon^4 \delta^6 c^2 - 
\frac{223729}{72} \epsilon^4 \delta^8 a d c^2  - \nonumber \\
&& \frac{7095}{16} \epsilon^3 \delta^6 c^2 + 
\frac{223729}{36} \epsilon^4 \delta^8 a^2 d^2 + \frac{15609}{128} \epsilon^5 \delta^7 \theta^4
 z^2 -
 \frac{5203}{64} \epsilon^5 \delta^7 \theta^4 + \nonumber \\
&& \frac{7095}{4} \epsilon^3 \delta^6 a d 
- \frac{5203}{4} \epsilon^4 \delta^6 a d \bigg ) \bigg \}
\end{eqnarray}
\par
Let now the extended tanh equation (\ref{eth}) be used as simplest equation.
The balance equation is the same as in the case when the equation of Riccati is used
as simplest equation. The application of the modified method of scimplest equation
leads to a system of nonlinear algebraic relationships among the parameters of the
equations and the parameters of the solution. One solution of this system is
\begin{eqnarray}\label{sol_sys_eth2}
\theta_2 &=& -6 \frac{\overline{a}^4 (p_5 + 2 p_6)}{p_4}; \hskip.5cm
 \theta_1 =0 \nonumber \\
\theta_0 &=& \frac{-p_2p _5 - 2 p_2 p_6 + 24 \overline{a}^2 \overline{d}^2 
p_5 p_6 + 16 \overline{a}^2 \overline{d}^2 p_6^2 + 8 \overline{a}^2 \overline{d}^2
 p_5^2 + 2 p_3 p_4
}{2 p_4 (p_5 + p_6)} \nonumber \\
v &=& \frac{V}{4 p_1 p_4 (p_5 + p_6)^2} \nonumber \\
V&=&  -2 p_2^2 p_5 p_6 - p_2^2 p_5^2 + 16 p_5^4 d^4 a^4 + 4 p_3^2 p_4^2 + 80 p_5^2 \overline
{d}^4 \overline{a}^4 p_6^2 + 64 p_5^3 \overline{d}^4 \overline{a}^4
 p_6 + \nonumber \\
&& 32 p_5 \overline{d}^4 \overline{a}^4 p_6^3 - 4 p_2 p_3 p_4 p_6 \nonumber \\
\end{eqnarray}
and the solution of Eq. (\ref{gen_kdv}) is
\begin{eqnarray}\label{sol_gen_kdv_eth}
\eta (\xi)  &=& \frac{-p_2p _5 - 2 p_2 p_6 + 24 \overline{a}^2 \overline{d}^2 p_5 p_6 + 16 
\overline{a}^2 \overline{d}^2 p_6^2 + 8 \overline{a}^2 \overline{d}^2
 p_5^2 + 2 p_3 p_4
}{2 p_4 (p_5 + p_6)} - \nonumber \\
&&  6 \frac{\overline{a}^2 \overline{d}^2 (p_5 + 2 p_6)}{p_4}
\tanh^2[\overline{a} \overline{d} (\xi + \xi_0)]
\end{eqnarray}
The surface wave (\ref{sol_ch2}) corresponding to the solution (\ref{sol_gen_kdv_eth})
is
\begin{eqnarray}\label{kdv_wave3}
u(\xi) &\approx& \frac{1}{9801 \epsilon^2} \big[ \overline{a}^4 \overline{d}^4 \delta^4 
\epsilon^2 (-561924 \epsilon  - 2013561   + 842886 \epsilon  z^2) [\tanh( \overline{a} 
\overline{d} (\xi + \xi_0))]^4 + \nonumber \\
&& \overline{a}^2 \overline{d}^2 \delta^2 \epsilon (- 280962 \epsilon - 
2684748 a^2 d^2 \epsilon \delta^2 + 383130  + 749232 \epsilon^2 \delta^2 a^2 d^2 - \nonumber \\
&& 1123848 \epsilon^2 \delta^2 a^2 d^2 z^2) [\tanh( \overline{a} \overline{d} (\xi + 
\xi_0))]^2 + 26730 \epsilon - 
187308 a^2 d^2 \epsilon^2 \delta^2 - \nonumber \\
&& 894916 a^4 d^4 \epsilon^2 \delta^4 - 18225 + 255420 a^2 d^2 \epsilon \delta^2 - 187308 
\epsilon^3 \delta^4 a^4 d^4 + \nonumber \\
&& 280962 \epsilon^3 \delta^4 a^4 d^4 z^2 \big]
\end{eqnarray}
\par
For a further demonstration of the modified method of simplest equations
let us now study what solutions of the classic Kortweg - de Vries equation can be obtained
by this method. The classic Kortweg - de Vries equation 
is particular case of Eq. (\ref{gen_kdv}) for the case $p_4 = p_5 = p_6 =0$. In other words 
we shall discuss the equation
\begin{equation}\label{classic_kdv}
p_1 \frac{\partial \eta}{\partial \tau} + p_2 \eta \frac{\partial \eta}{\partial \zeta} + 
p_3 \frac{\partial^3 \eta}{\partial \zeta^3} =0
\end{equation}
which after the subsitution of $p_{1,2,3}$ from Eq. (\ref{parameters}) is reduced
to the Korteweg - de Vries equation. The balance equations for Eq. (\ref{classic_kdv})
are the same as for the Eq. (\ref{gen_kdv}). Lt us first discuss the case where the
equation of Bernoulli is used as simplest equation. The application of the modified
method of simplest equation reduces (\ref{classic_kdv}) to a system of 5 nonlinear algebraic
relationships among the parameters of the equation and parameters of the solution (\ref
{solution}) (we remember that $\nu =0$ and $V(\xi))$ is a solution of the Bernoulli equation).
A solution of this nonlinear algebraic system is
\begin{eqnarray}\label{classic_syssol1}
\theta_2 = -12 \frac{p_3 c^2}{p_2}; \hskip.5cm \theta_1 = -12 \frac{p_3 c a}{p_2};\hskip.5cm
\theta_0 = \frac{p_1 v - p_3 a^2}{p_2}
\end{eqnarray}
The corresponding solution of (\ref{classic_kdv})  for $c<0$ and $a>0$ is
\begin{eqnarray}\label{classic_sol1}
\eta = \frac{p_1 v - p_3 a^2}{p_2} - 12 \frac{p_3 c a^2}{p_2} \bigg [
\frac{\exp(a [\xi + \xi_0)]}{1 - c \exp [a(\xi + \xi_0)]}  \bigg ]
\bigg \{ 1 + c \bigg [
\frac{\exp(a [\xi + \xi_0)]}{1 - c \exp [a(\xi + \xi_0)]}  \bigg ] \bigg \}
\nonumber \\
\end{eqnarray}
For the case $a<0$; $c>0$ the solution of (\ref{classic_kdv}) is
\begin{eqnarray}\label{classic_sol2}
\eta = \frac{p_1 v - p_3 a^2}{p_2} + 12 \frac{p_3 c a^2}{p_2} \bigg [
\frac{\exp(a [\xi + \xi_0)]}{1 + c \exp [a(\xi + \xi_0)]}  \bigg ]
\bigg \{ 1 - c \bigg [
\frac{\exp(a [\xi + \xi_0)]}{1 + c \exp [a(\xi + \xi_0)]}  \bigg ] \bigg \}
\nonumber \\
\end{eqnarray}
By setting $p_1 =2$; $p_2 =3$ and $p_3 = \frac{1}{3} \delta^2$ in Eqs. (\ref{classic_sol1})
and (\ref{classic_sol2}) we obtain the corresponding solutions of the classic
Korteweg - de Vries equation.
\par
Let now the equation of Riccati (\ref{riccati}) is used as simplest equation. For this
case the application of the modified method of simplest equation leads to a system of
$6$ nonlinear relationships among the parameters of the equation and parameters of the
solution. One solution of this system is
\begin{equation}\label{classic_syssol2}
\theta_2 = -12 \frac{p_3 a^2}{p_2}; \hskip.5cm \theta_1 = -12 \frac{p_3 a c }{p_2};
\hskip.5cm \theta_0 = \frac{p_1 v - 8 p_3 a d - p_3 c^2}{p_2}
\end{equation}
The solution of Eq. (\ref{classic_kdv}) based on the solution (\ref{sol_ric1}) of the
Riccati equation is 
\begin{eqnarray}\label{classic_sol3}
\eta &=& \frac{p_1 v - 8 p_3 a d - p_3 c^2}{p_2} + \frac{p_3 a c }{p_2} \bigg \{
\frac{c}{2a} + \frac{\theta}{2 a} \tanh \left [ \frac{\theta (\xi + \xi_0)}{2} \right ]\bigg \}
\bigg \{ 1 - \nonumber \\
&& \frac{a}{c} \bigg \{
\frac{c}{2a} + \frac{\theta}{2 a} \tanh \left [ \frac{\theta (\xi + \xi_0)}{2} \right ]\bigg 
\} \bigg \}
\end{eqnarray}
and the solution of Eq. (\ref{classic_kdv}) based on the solution (\ref{sol_ric2}) of the
Riccati equation is 
\begin{eqnarray}\label{classic_sol4}
\eta = \frac{p_1 v - 8 p_3 a d - p_3 c^2}{p_2} -12 \frac{p_3 a c }{p_2} \bigg \{
- \frac{c}{2a} - \frac{\theta}{2a} \tanh
\left(\frac{\theta (\xi + \xi_0)}{2} \right) + \nonumber \\
\cfrac{\exp \left(\cfrac{\theta (\xi+\xi_0)}{2}
\right)}{ 2 \cosh \left(\cfrac{\theta (\xi +\xi_0)}{2} \right) \left[\cfrac{a}{\theta}+
2 C^{*} \exp \left(\cfrac{\theta (\xi + \xi_0)}{2} \right) \cosh \left(\cfrac{\theta
(\xi + \xi_0)}{2} \right)  \right]} \bigg \} + \nonumber \\
 \bigg \{  1+ \frac{a}{c} \bigg[ - \frac{c}{2a} - \frac{\theta}{2a} \tanh
\left(\frac{\theta (\xi + \xi_0)}{2} \right) + \nonumber \\
\cfrac{\exp \left(\cfrac{\theta (\xi+\xi_0)}{2}
\right)}{ 2 \cosh \left(\cfrac{\theta (\xi +\xi_0)}{2} \right) \left[\cfrac{a}{\theta}+
2 C^{*} \exp \left(\cfrac{\theta (\xi + \xi_0)}{2} \right) \cosh \left(\cfrac{\theta
(\xi + \xi_0)}{2} \right)  \right]} \bigg ] \bigg \} \nonumber \\
\end{eqnarray}
From Eqs. (\ref{classic_sol3}) and (\ref{classic_sol4}) by setting $p_1 =2$; $p_2 =3$ and 
$p_3 = \frac{1}{3} \delta^2$ one obtains the corresponding solutions of the classic
Korteweg - de Vries equation.
\par
Finally the the extended tahn equation (\ref{eth}) be used as a simplest equation. The 
balance equation for this case is the same as in the case when the simplest equation is the 
equation of Riccati. The application of the modified method of simplest equation
leads to a system of 6 nonlinear algebraic equations for the parameters of the solution.
One solution of this system is
\begin{eqnarray}\label{sol_syst_eth1}
\theta_2 = - 12 \frac{p_3 \overline{a}^4}{p_2}; \hskip.5cm \theta_1 = 0; \hskip.5cm
\theta_0 = \frac{v p_1 + 8 p_3 \overline{a}^2 \overline{d}^2}{p_2}
\end{eqnarray}
The sollution of (\ref{classic_kdv}) is
\begin{eqnarray}\label{sol_kdv_eth}
\eta (\xi)  =  \frac{v p_1 + 8 p_3 \overline{a}^2 \overline{d}^2}{p_2}
- 12 \frac{p_3 \overline{a}^4}{p_2} \frac{\overline{d}^2}{\overline{a}^2}
\tanh^2 [\overline{a} \overline{d} (\xi + \xi_0)]
\end{eqnarray}
\section{Exact traveling - wave solutions of the generalized Camassa - Holm equation}
Let first the equation of Bernoulli be used as simplest equation
and the values of parameters from (\ref{bal_ch1}) are $J=3$,
$I=2$, $k=2$. Then from (\ref{bal_ch1}) one obtains $\nu_1=2$.
The equation we shall solve is obtained from (\ref{gen_ch}) when
$I=2$, $J=3$. This equation is
\begin{eqnarray}\label{gen_chx1}
\frac{\partial U}{\partial T} + p_1 \frac{\partial U}{\partial X} - 
p_2 \frac{\partial^3 U}{\partial X^2 \partial T} - p_3 \frac{\partial^3}{\partial X^3} 
(q_0 + q_1 U + q_2 U^2) + \nonumber \\
p_4 \frac{\partial}{\partial X} \left[ \frac{1}{2} (r_0 + r_1 U + r_2 U^2 + r_3 U^3) 
+ p_5 q_2 \left( \frac{\partial U}{\partial X}\right)^2 \right ] =0
\end{eqnarray}
The solution of Eq. (\ref{gen_ch1}) is searched in the form
\begin{equation}\label{s1}
U(\xi) = \theta_0 + \theta_1 V(\xi) + \theta_2 V(\xi)^2
\end{equation}
where $V(\xi)$ is solution of the equation of Bernoulli (\ref{bernoulli})
for $k=2$. The substitution of (\ref{s1}) in (\ref{gen_ch1}) leads to
a large system of $7$ nonlinear relationships among the parameters of the
generalized Camassa - Holm equation and parameters of the solution (\ref{s1})
One solution of this system of equations is
\begin{eqnarray}\label{ch_sys_sol1}
\theta_2 &=& 8 \frac{q_2 a c (5 p_3 - p_4 p_5)}{p_4 r_3}; \hskip.5cm
\theta_1 = 8 \frac{q_2 c^2 (5 p_3 - p_4 p_5)}{p_4 r_3} \nonumber \\
\theta_0 &=& \frac{\Theta_0}{6 q_2 p_4 r_3 (4 p_3 - p_4 p_5)} \hskip.5cm
r_1 = \frac{R_1}{12 p_2 p_4 r_3 (4 p_3 - p_4 p_5)} \nonumber \\
\Theta_0 &=& - 36 q_2^2 a^2 p_4 p_5 p_3 + 4 q_2^2 a^2 p_4^2 p_5^2 - 3 p_2 v p_4 r_3 + 3 
p_3 q_1 p_4 r_3 + 80 q_2^2 a^2 p_3^2 - \nonumber \\
&& 10 q_2 p_3 r_2 p_4 + 2 q_2 p_4^2 p_5 r_2 \nonumber \\
R_1 &=& - ( 1520 q_2^4 a^4 p_4^2 p_5^2 p_3^2 - 256 q_2^4 a^4 p_4^3 p_5^3 p_3 - 4 q_2^2 p_4^4 
p_5^2 r_2^2 + 32 q_2^2 p_3 r_2^2 p_4^3 p_5 - \nonumber \\
&& 60 q_2^2 p_3^2 r_2^2 p_4^2 + 3840 q_2^4 a^4 p_3^4 - 12 p_3^2 q_1 p_4^2 r_3 q_2 r_2 + 12 
p_2 v p_4^2 r_3 q_2 p_3 r_2 - \nonumber \\
&& 18 p_2 v p_4^2 r_3^2 p_3 q_1 + 9 p_3^2 q_1^2 p_4^2 r_3^2 + 16 q_2^4 a^4 p_4^4 p_5^4 +9 
p_2^2 v^2 p_4^2 r_3^2 - \nonumber \\
&& 3968 q_2^4 a^4 p_4 p_5 p_3^3 - 24 v q_2^2 r_3 p_4^3 p_5^2 + 192 v q_2^2 r_3 p_4^2 p_3 p_5 
- 384 v q_2^2 r_3 p_4 p_3^2 - 
\nonumber \\
&& 192 p_1 q_2^2 r_3 p_4^2 p_3 p_5 + 384 p_1 q_2^2 r_3 p_4 p_3^2 + 24 p_1 q_2^2 r_3 
p_4^3 p_5^2) \nonumber \\
\end{eqnarray}
The corresponding solution of Eq. (\ref{gen_chx1}) is 
\begin{eqnarray}\label{sx1}
U(\xi) = \frac{\Theta_0}{6 q_2 p_4 r_3 (4 p_3 - p_4 p_5)} + 8 \frac{q_2 c^2 (5 p_3 - p_4 
p_5)}{p_4 r_3} \left \{ \frac{a \exp [a (\xi + \xi_0)]}{1 - c \exp [a (\xi + \xi_0)]} \right \}
+ \nonumber \\
8 \frac{q_2 a c (5 p_3 - p_4 p_5)}{p_4 r_3} \left \{ \frac{a \exp [a (\xi + \xi_0)]}{1 - c 
\exp [a (\xi + \xi_0)]} \right \}^2 \nonumber \\
\end{eqnarray}
for $c<0$ and $a >0$. For $c>0$ and
$a <0$ the solution of Eq. (\ref{gen_chx1}) is
\begin{eqnarray}\label{sx2}
U(\xi) = \frac{\Theta_0}{6 q_2 p_4 r_3 (4 p_3 - p_4 p_5)} - 8 \frac{q_2 c^2 (5 p_3 - p_4 
p_5)}{p_4 r_3} \left \{ \frac{a \exp [a (\xi + \xi_0)]}{1 + c \exp [a (\xi + \xi_0)]} \right \}
+ \nonumber \\
8 \frac{q_2 a c (5 p_3 - p_4 p_5)}{p_4 r_3} \left \{ \frac{a \exp [a (\xi + \xi_0)]}{1 + c 
\exp [a (\xi + \xi_0)]} \right \}^2 \nonumber \\
\end{eqnarray}
\par
Now let the equation of Riccati (\ref{riccati}) be used as simplest equation. From the
possible balance equations Eq. (\ref{bal_ch2}) will be used below with values of the
parameters $J=1$; $\nu_1 =2$; $I=2$. Thus the equation which will be solved is 
\begin{eqnarray}\label{gen_chx2} 
\frac{\partial U}{\partial T} + p_1 \frac{\partial U}{\partial X} - 
p_2 \frac{\partial^3 U}{\partial X^2 \partial T} - p_3 \frac{\partial^3}{\partial X^3} 
(q_0 + q_1 U + q_2 U^2) + \nonumber \\
p_4 \frac{\partial}{\partial X} \left[ \frac{1}{2} (r_0 + r_1 U) 
+ p_5 q_2 \left( \frac{\partial U}{\partial X}\right)^2 \right ] =0
\end{eqnarray}
In this case the application of the method of simplest equation to Eq. (\ref{gen_chx2}) 
leads to a system of 8 nonlinear algebraic relationships among the parameters
of the solved equation and the parameters of the solution. One solution of this
system is 
\begin{eqnarray}\label{ch_sys_sol2}
\theta_2 &=& - 15 \frac{a^2 (-p_4 r_1 + 2 v - 2 p_1)}{q_2 p_4 p_4 (-8 c^2 d a + 16 a^2 d^2 + 
c^4)} \nonumber \\
\theta_1 &=&  - 15 \frac{a c (-p_4 r_1 + 2 v - 2 p_1)}{q_2 p_4 p_4 (-8 c^2 d a + 16 a^2 d^2 + 
c^4)} \nonumber \\
\theta_0 &=& \frac{\Theta_0}{q_2 p_4 p_5 (-8 c^2 d a + 16 a^2 d^2 + c^4)} \nonumber \\
\Theta_0 &=& \frac{1}{4} (40 a d p_4 r_1 - 80 a d v + 80 a d p_1 + 5 c^2 p_4 r_1 - 10 c^2 v + 
10 c^2 p_1 + \nonumber \\
&& 16 p_4 p_5 q_1 c^2 d a - 32 p_4 p_5 q_1 a^2 d^2 - 2 p_4 p_5 q_1 c^4 - 80 p_2 v c^2 d a + 
160 p_2 v a^2 d^2 + \nonumber \\
&& 10 p_2 v c^4) \nonumber \\
\end{eqnarray}
The corresponding solution of Eq. (\ref{gen_chx2}) based on the solution (\ref{sol_ric1})
of the Riccati equation is
\begin{eqnarray}\label{sx3}
U(\xi) &=& \frac{\Theta_0}{q_2 p_4 p_5 (-8 c^2 d a + 16 a^2 d^2 + c^4)} + \nonumber \\
&& 15 \frac{a c (-p_4 r_1 + 2 v - 2 p_1)}{q_2 p_4 p_4 (-8 c^2 d a + 16 a^2 d^2 + 
c^4)} \bigg[ \frac{c}{2 a} + \frac{\theta}{2 a} \tanh \bigg( \frac{\theta (\xi + \xi_0)}{2}
\bigg)\bigg] - \nonumber \\
&& 15 \frac{a^2 (-p_4 r_1 + 2 v - 2 p_1)}{q_2 p_4 p_4 (-8 c^2 d a + 16 a^2 d^2 + 
c^4)} \bigg[ \frac{c}{2 a} + \frac{\theta}{2 a} \tanh \bigg( \frac{\theta (\xi + \xi_0)}{2}
\bigg)\bigg]^2 \nonumber \\
\end{eqnarray}
where $\theta^2 = c^2 - 4 a d >0$.
\par
The solution of (\ref{gen_chx2}) based on the solution (\ref{sol_ric2}) of the Riccati
equation is
\begin{eqnarray}\label{sx4}
U(\xi) &=& \frac{\Theta_0}{q_2 p_4 p_5 (-8 c^2 d a + 16 a^2 d^2 + c^4)} - \nonumber \\
&& 15 \frac{a c (-p_4 r_1 + 2 v - 2 p_1)}{q_2 p_4 p_4 (-8 c^2 d a + 16 a^2 d^2 + 
c^4)} \bigg \{  - \frac{c}{2a} - \frac{\theta}{2a} \tanh
\left(\frac{\theta (\xi + \xi_0)}{2} \right) + \nonumber \\
&& \cfrac{\exp \left(\cfrac{\theta (\xi+\xi_0)}{2}
\right)}{ 2 \cosh \left(\cfrac{\theta (\xi +\xi_0)}{2} \right) \left[\cfrac{a}{\theta}+
2 C^{*} \exp \left(\cfrac{\theta (\xi + \xi_0)}{2} \right) \cosh \left(\cfrac{\theta
(\xi + \xi_0)}{2} \right)  \right]} \bigg \}  - \nonumber \\
&&  15 \frac{a^2 (-p_4 r_1 + 2 v - 2 p_1)}{q_2 p_4 p_4 (-8 c^2 d a + 16 a^2 d^2 + 
c^4)} \bigg \{  - \frac{c}{2a} - \frac{\theta}{2a} \tanh
\left(\frac{\theta (\xi + \xi_0)}{2} \right) + \nonumber \\
&& \cfrac{\exp \left(\cfrac{\theta (\xi+\xi_0)}{2}
\right)}{ 2 \cosh \left(\cfrac{\theta (\xi +\xi_0)}{2} \right) \left[\cfrac{a}{\theta}+
2 C^{*} \exp \left(\cfrac{\theta (\xi + \xi_0)}{2} \right) \cosh \left(\cfrac{\theta
(\xi + \xi_0)}{2} \right)  \right]} \bigg \}^2 \nonumber \\
\end{eqnarray}
\par
Let now the extended tanh equation (\ref{eth}) be the simplest equation. Let
in addition $I = J=2$ and $\nu_1 =2$.  The equation which is solved for this case is
\begin{eqnarray}\label{gen_chx4} 
\frac{\partial U}{\partial T} + p_1 \frac{\partial U}{\partial X} - 
p_2 \frac{\partial^3 U}{\partial X^2 \partial T} - p_3 \frac{\partial^3}{\partial X^3} 
(q_0 + q_1 U + q_2 U^2) + \nonumber \\
p_4 \frac{\partial}{\partial X} \left[ \frac{1}{2} (r_0 + r_1 U + r_2 U^2) 
+ p_5 q_2 \left( \frac{\partial U}{\partial X}\right)^2 \right ] =0
\end{eqnarray}
The application of the modified method of simplest equation leads to a system of
8 noninear algebraic relationships among the parameters of the solution and para
meters of the solved equation. One solution of this system is
\begin{eqnarray}\label{ss1}
\theta_2 &=& 60 \overline{a}^4\frac{( - 2 v p_5 q_2 + 2 p_1 p_5 q_2 + r_1 p_4 p_5 q_2 + 5 r_2 
p_2 v - r_2 p_4 p_5 q_1)
}{p_4 (64 p_5^2 q_2^2 \overline{d}^4 \overline{a}^2 - 25 r_2^2)} \nonumber \\
\theta_1 &=& 0 \nonumber \\
\theta_0 &=& \frac{\Theta_0}{p_4 (-25 r_2^2 + 64 p_5^2 q_2^2 \overline{d}^4
 \overline{a}^4)}  \nonumber \\
\Theta_0 &=& \frac{1}{2} (160 p_5 q_2 \overline{a}^2 \overline{d}^2 v - 
160 p_5 q_2 \overline{a}^2 \overline{d}^2 p_1 - 80 p_5 q_2 \overline{a}^2 
\overline{d}^2 r_1 p_4 - 400 \overline{a}^2 \overline{d}^2 r_2 p_2 v + \nonumber \\
&& 80 p_5 \overline{a}^2 \overline{d}^2 r_2 p_4 q_1 + 320 p_2 v p_5 q_2 \overline{d}^4 \overline{a}^4 - 50 r_2 v + 50 r_2 p_1 + 25 r_2 r_1 p_4 - \nonumber \\
&& 64 p_4 p_5^2 q_1 q_2 \overline{d}^4 \overline{a}^4) \nonumber \\
p_3 &=& \frac{1}{5} p_4 p_5 \nonumber \\
\end{eqnarray}
The corresponding solution is as follows
\begin{eqnarray}\label{ss_sol}
U ( \xi )  &=& \frac{\Theta_0}{p_4 (-25 r_2^2 + 64 p_5^2 q_2^2 \overline{d}^4
 \overline{a}^4)} + \nonumber \\
&& 60 \overline{a}^2 \overline{d}^2 \frac{( - 2 v p_5 q_2 + 2 p_1 p_5 q_2 + r_1 p_4 p_5 q_2 + 5 r_2 p_2 v - r_2 p_4 p_5 q_1)
}{p_4 (64 p_5^2 q_2^2 \overline{d}^4 \overline{a}^2 - 25 r_2^2)} 
\tanh^2[\overline{a} \overline{d} (\xi + \xi_0)]
\nonumber \\
\end{eqnarray}

\par
Let us now discuss the equation
\begin{equation}\label{almost_ch}
\frac{\partial U}{\partial T} + p_1 \frac{\partial U}{\partial X} - 
p_2 \frac{\partial^3 U}{\partial X^2 \partial T} + 3 p_4 U \frac{\partial U}{\partial X}
-2 p_3 \frac{\partial U}{\partial X} \frac{\partial^2 U}{\partial X^2}
- p_5 U \frac{\partial^3 U}{\partial X^3} =0
\end{equation}
which can be reduced to Eq. (\ref{gen_ch1}) and then to the Camassa - Holm equation when 
$p_5 = p_3$. If the equation of Bernoulli (\ref{bernoulli}) is used as simplest equation
in the modified method of simplest equation we obtain balance equation of the kind
$2 \nu_1 + 3 (k-1) = 2 \nu_1 + 3(k-1)$. The simplest case is $\nu_1 = 2$, $k=2$.
For this case the application of the modified method of simplest equation reduces
Eq. (\ref{almost_ch}) to a system of 7 nonlinear algebraic relationships among the
parsmeters of the equation and the parameters of the solution. One solution of
this system is
\begin{eqnarray}\label{sol_syst_almost_ch}
p_5 &=& - p_3; \hskip.5cm v = \frac{12 p_1 c^2 p_3 - 9 p_4^2 \theta_2 + p_3^2 \theta_2 
a^4}{12 c^2 (p_3 + 3 p_4 p_2)} \nonumber \\
\theta_0 &=& - \frac{ 3 p_4 \theta_2 - p_3 \theta_2 a^2 - 3 \theta_2 a^2 p_4 p_2 + 12 p_2 p_1 
c^2 + p_2 p_3 \theta_2 a^4}{12 c^2 (p_3 + 3 p_4 p_2)} \nonumber \\
\theta_1 &=& \frac{\theta_2 a}{c}
\end{eqnarray}
the corresponding solution of Eq. (\ref{almost_ch}) is 
\begin{eqnarray}\label{sol1_almost_ch}
U(\xi) = - \frac{ 3 p_4 \theta_2 - p_3 \theta_2 a^2 - 3 \theta_2 a^2 p_4 p_2 + 12 p_2 p_1 
c^2 + p_2 p_3 \theta_2 a^4}{12 c^2 (p_3 + 3 p_4 p_2)} + \nonumber \\
\frac{\theta_2 a}{c}
\left \{ \frac{a \exp [a (\xi + \xi_0)]}{1 - c \exp[a (\xi + \xi_0)]} \right \}
+ \theta_2 \left \{ \frac{a \exp [a (\xi + \xi_0)]}{1 - c \exp[a (\xi + \xi_0)]} \right \}^2
\end{eqnarray}
for the case $c<0$, $a>0$. For the case $c>0$, $a<0$ the solution is
\begin{eqnarray}\label{sol2_almost_ch}
U(\xi) = - \frac{ 3 p_4 \theta_2 - p_3 \theta_2 a^2 - 3 \theta_2 a^2 p_4 p_2 + 12 p_2 p_1 
c^2 + p_2 p_3 \theta_2 a^4}{12 c^2 (p_3 + 3 p_4 p_2)} - \nonumber \\
\frac{\theta_2 a}{c}
\left \{ \frac{a \exp [a (\xi + \xi_0)]}{1 + c \exp[a (\xi + \xi_0)]} \right \}
+ \theta_2 \left \{ \frac{a \exp [a (\xi + \xi_0)]}{1 + c \exp[a (\xi + \xi_0)]} \right \}^2
\end{eqnarray}
It must be noted that because of the requirement $p_5 = -p_3$ the above two solutions
can not be reduced to solutions of the Camassa - Holm equation. If the rquirement
was $p_5 = p_3$ such reduction would be possible. However the requirement is  
$p_5 = -p_3$ and the reduction is not possible.
\section{Concluding remarks}
In this paper we have studied two nonlinear partial differential equations connected 
to the water waves: extended Korteweg - de Vries equation and generalized 
Camassa - Holm equation. The goal of the study was to obtain exact traveling-wave solutions 
of these equations.  This goals was achieved on the basis of the  method of 
simplest equation which is an effective tool for obtaining exact analytical solutions
for large classes of nonlinear partial differential equations. We have applied a 
modification of this method based on use of balance equation. The equations of Bernoulli, 
Riccati, and extended tanh - equation have been used as simplest equations. 
We have obtained numerous exact traveling-wave solutions of the studied equations and
for the case of the extended Korteweg - de Vries equation we have obtained three expressions 
for surface water waves which correspond to the exact solutions of the extended Korteweg -
de Vries equation. As a side result we have applied the methodology to the classic
Korteweg - de Vries equation and we have demonstrated that by means of the modified 
method of simplest equation exact solutions of this classic nonlinear partial
differential equation can be obtained too.
\par
This paper is manuscript Nr. AMC-D-10-04363 submitted to {\sl Applied Mathematics and
Computation} on 27th of December 2010. The manuscript is still under review.


\begin{thebibliography}{99}
\bibitem{c1}
	J. C. Tannenhill, D. A. Anderson, R. H. Pletcher. Computational fluid mechanics and 
	heat transfer. Taylor and Francis, Philadelphia, PA, 1997.
\bibitem{c2}
	F. Dias. Nonlinear gravity and capillary-gravity waves. Annual Review of Fluid Mechanics
	31 (1999) 301 - 346.
\bibitem{c3}
	N. K. Vitanov, F. H. Busse. Bounds on the heat transport in a horizontal fluid layer with
	stress-free boundaries. ZAMP 48 (1997) 310 - 324.
\bibitem{scott}
	A. C. Scott. Nonlinear  science. Emergence and dynamics of coherent
	structures. Oxford University Press, Oxford, UK, 1999.
\bibitem{c4}
	N. K. Vitanov. Upper bound on the heat transport in a horizontal fluid layer of infinite
	Prandtl number. Phys. Lett. A 248 (1998)  338 - 346.
\bibitem{c5}
	T. B. Benjamin, J. L. Bona, J. J. Mahoni. Model equations for long waves in nonlinear
	dispersive systems. Phil. Trans. Roy. Soc. London A 272 (1972) 47 - 78.
\bibitem{c6}
	J. W. Miles. Solitary waves. Annual Review of Fluid Mechanics 12 (1980) 11 - 43.	
\bibitem{c7}
	H. Kantz, D. Holstein, M. Ragwitz, N. K. Vitanov. Markov chain model for turbulent wind 
	speed data. Physica A  342 (2004) 315 - 321. 
\bibitem{c8}
	N. K. Vitanov. Upper bound on the heat transport in a layer of fluid of infinite 
	Prandtl number, rigid lower boundary, and stress-free upper boundary. Phys. Rev. E 
	61 (2000) 956 - 959.
\bibitem{scot1}
	A. Scott. Neuroscience: A mathematical primer.  Springer, New York, 2002.
\bibitem{longtin}
	A. Longtin, J. G. Milton. Modeling autonomous oscillations in the
	human pupil light reflex using nonlinear delay-differential equations.
	Bull. Math. Biology  51 (1989) 605 - 624.
\bibitem{ac}
	M. Ablowitz, P. A. Clarkson. Solitons, nonlinear evolution equations
	and inverse scattering. Cambridge University Press, Cambridge, UK, 1991.
\bibitem{wazwaz1}
	A. -M. Wazwaz. Partial differential equations and solitary waves theory. Springer, 
	Berlin, 2009.
\bibitem{infeld}
	E. Infeld, G. Rowlands. Nonlinear waves, solitons and chaos. Cambridge University Press, 
	Cambridge, UK, 1990.
\bibitem{newell}
	A. C. Newell. Solitons in mathematics and physics. SIAM, Philadelphia, PA, 1985. 
\bibitem{murr}
	J. D. Murray. Lectures on nonlinear differential equation models
	in biology. Oxford University Press, Oxford, England, 1977.
\bibitem{perko}
	L. Perko. Differential equations and dynamical systems. Springer,
	New York, 2001.
\bibitem{may}
	R. M. May. Stability and complexity in model ecosystems. Princeton
	University Press, New Jersey, 2001.
\bibitem{temam}
	R. Temam. Navier - Stokes equations: Theory and numerical analysis. 
	AMS Chelsea Publishing, Providence, R. I., 2001.
\bibitem{foias}
	C. Foias, O. Manley, R. Rosa, R. Temam. Navier - Stokes equations and
	turbulence. Cambridge University Press, Cambridge, UK, 2001.
\bibitem{holmes1}
	P. Holmes, J. L. Lumley, G. Berkooz. Turbulence, coherent structures, dynamical systems 
	and symmetry. Cambridge University Press, Cambridge, UK, 1996.
\bibitem{ott1}
	N. P. Hoffmann, N. K. Vitanov. Upper bounds on energy dissipation in 
	Couette-Ekman flow. Phys. Lett. A 255 (1999) 277 - 286. 
\bibitem{ott2}
	N. K. Vitanov. Upper bounds on the heat transport in a porous layer. Physica D
	136 (2000) 322 - 339.
\bibitem{boeck}
	T. Boeck, N. K. Vitanov. Low-dimensional chaos in zero-Prandtl-number 
	Benard-Marangoni convection. Phys. Rev. E 65 (2002) Article number
	037203. 
\bibitem{epjb1}
	N. K. Vitanov. Convective heat transport in a fluid layer of infinite Prandtl number:
	upper bounds for the case of rigid lower boundary and stress - free upper boundary.
	European Physical Journal B  15 (2000) 349 - 355.
\bibitem{vx}
	N. K. Vitanov. Upper bounds on convective heat transport in a rotating layer of infinite
	Prandtl number: Case of intermediate taylor numbers. Phys. Rev. E 62 (2000) 3581 - 3591.
\bibitem{gardner}
	C. S. Gardner, J. M. Greene, M. D. Kruskal, R. R. Miura. Method for solving
	Korteweg- de Vries equation. Phys. Rev. Lett. 19 (1967) 1095 - 1097.
\bibitem{ablowitz1}
	M. J. Ablowitz, D. J. Kaup, A. C. Newell. Nonlinear evolution equations of
	physical significance. Phys. Rev. Lett. 31 (1973) 125 - 127.
\bibitem{remos}
	M. Remoissenet. Waves called solitons. Springer, Berlin; 1993.
\bibitem{ablowitz2}
	M. J. Ablowitz, D. J. Kaup, A. C. Newell, H. Segur. Inverse scattering
	transform - Fourier analysis for nonlinear problems. Studies in
	Applied Mathematics 53 (1974) 249 - 315.
\bibitem{hirota}
	R. Hirota. Exact solution of Korteweg-de Vries equation for
	multiple collisions of solitons. Phys. Rev. Lett. 27 (1971) 1192 - 1194.
\bibitem{kudr90}
	N. A. Kudryashov. Exact solutions of the generalized Kuramoto -
	Sivashinsky equation. Phys. Lett. A 147 (1990) 287 - 291.
\bibitem{yan}
	Z. Y. Yan. New explicit travelling wave solutions for two new
	integrable coupled nonlinear evolution equations. Phys. Lett. A 292 (2001)
	100 - 106.
\bibitem{fan00}
	E. G. Fan. Extended tanh-function method and its application to
	nonlinear equations. Phys. Lett. A 277 (2000) 212 - 218.
\bibitem{fan02}
	E. G. Fan. Traveling wave solutions for nonlinear equations using symbolic computation.
	Computers \& Mathematics with Applications  43 (2002) 671 - 680.
\bibitem{hex}
	J. -H. He, X,-H. Wu. Exp-function method for nonlinear wave equations. Chaos Solitons \& 
	Fractals 30 (2006) 700 - 708.
\bibitem{kudr05x}
	N. A. Kudryashov. Simplest equation method to look for exact solutions
	of nonlinear differential equations. Chaos Solitons \& Fractals
	24 (2005) 1217 - 1231.
\bibitem{vdk10}
	N. K. Vitanov, Z. I. Dimitrova, H. Kantz. Modified method of simplest equation
	and its application to nonlinear PDEs. Applied Mathematics and Computation
	216 (2010) 2587 - 2595.
\bibitem{vit11}
	N. K. Vitanov. Modified method of simplest equation: Powerful tool for obtaining exact and
	approximate travelling-wave solutions of nonlinear PDEs. Communications in Nonlinear 
	Science and Numerical Simulation 16 (2011) 1176 - 1185.
\bibitem{zeppet}
	M. J. Ablowitz, A. Zeppetela. Explicit solutions of Fisher equation
	for a specifical wave speed. Bull. Math. Biol. 41 (1979) 835 - 840.
\bibitem{dv1}
	Z. I. Dimitrova, N. K. Vitanov. Influence of adaptation on the nonlinear
	dynamics of a system of competing populations. Phys. Lett. A 272 (2000)
	368 - 380.
\bibitem{dv2}
	Z. I. Dimitrova, N. K. Vitanov. Dynamical consequences of adaptation of
	growth rates in a system of three competing populations. J. Phys. A:
	Math. Gen. 34 (2001) 7459 - 7473.
\bibitem{dv3}
	Z. I. Dimitrova, N. K. Vitanov. Adaptation and its impact on the dynamics
	of a system of three competing populations. Physica A 300 (2001) 91 -
	115.
\bibitem{dv4}
	Z. I. Dimitrova, N. K. Vitanov. Chaotic pairwise competition. Theoretical
	Population Biology 66 (2004) 1 - 12.
\bibitem{dv5}	 
	N. K. Vitanov, Z. I. Dimitrova, H. Kantz. On the trap of extinction and its
	elimination. Phys. Lett. A 349 (2006) 350 - 355.
\bibitem{vit09}
	N. K. Vitanov, I. P. Jordanov, Z. I. Dimitrova. On nonlinear dynamics of 
	interacting populations: Coupled kink waves in a system of two
	populations. Communications in Nonlinear Science and  Numerical Simulation 
	14 (2009) 2379 - 2388.
\bibitem{vit09a}
	N. K. Vitanov, I. P. Jordanov, Z. I. Dimitrova. On nonlinear population waves.
	Applied Mathematics and Computation. 215 (2009) 2950 - 2964.
\bibitem{wang}
	X. Y. Wang. Exact and explicit wave solutions for the generalized Fisher
	equation. Phys. Lett. A 131 (1988) 277 - 279.
\bibitem{kudr05}
	N. A. Kudryashov. Exact solitary waves of the Fisher equation. Phys.
    Lett. A  342 (2005) 99 - 106.
\bibitem{vd10a}
	N. K. Vitanov, Z. I. Dimitrova. Application of the method of simplest equation for 
	obtaining exact traveling-wave solutions for two classes of model PDEs from ecology and 
	population dynamics. Communications in Nonlinear Science and Numerical Simulation 15 (2010)
	2836 - 2845.
\bibitem{holmes94}
	E. E. Holmes, M. A. Lewis, J. E. Banks, R. R. Veit. Partial differential
	equation in ecology: Spatial interactions and population dynamics.
	Ecology 75 (1994) 17 - 29.
\bibitem{vd10b}
	N. K. Vitanov, Z. I. Dimitrova, M. Ausloos.  Verhulst - Lotka - Volterra (VLV) model of
	ideological struggle. Physica A  389 (2010) 4970 - 4980.
\bibitem{lou}
	S. Lou. Symmetry analysis and exact solutions of the 2+1 - dimensional
	sine - Gordon system. J. Math. Phys. 41 (2000) 6509 - 6524.
\bibitem{wazwaz}
	A. -M. Wazwaz. The tanh method: exact solutions of the sine - Gordon and
	the sh-Gordon equations. Applied Mathematics and Computation 167 (2005)
	1196 - 1210.
\bibitem{vit94a}
	N. K. Martinov, N. K. Vitanov. New class of running-wave solutions of the
	2+1-dimensional sine-Gordon equation. J. Phys. A: Math. Gen. 27 (1994)
	4611 - 4618.
\bibitem{vit96a}
	N. K. Vitanov. On traveling waves and double - periodic structures in
	two - dimensional sine - Gordon systems. J. Phys. A: Math. Gen. 29 (1996) 
	5195 - 5207.
\bibitem{clarkson} 
	P. A. Clarkson, E. L. Mansfield, A. E. Milne. Symmetries and exact solutions
	of a (2+1)-dimensional sine - Gordon system. Phil. Trans. Roy. 
	Soc. London A 354 (1996) 1807 - 1835.
\bibitem{vit96b}
	N. K. Vitanov, N. K. Martinov. On the solitary waves in the sine-Gordon
	model of the two-dimensional Josephson junction. Z. Phys. B 100 (1996)
    129 - 135.
\bibitem{vit98}
	N. K. Vitanov. Breather and soliton wave families for the sine - Gordon
	equation. Proc. Roy. Soc. London A 454 (1998) 2409 - 2423.
\bibitem{radha}
	R. Radha, M. Lakshamanan. The (2+1) - dimensional sine - Gordon
	equation; Integrability and localized solutions. J. Phys A: Math. Gen.
	29 (1996) 1551 - 1562.
\bibitem{nakamura}
	A. Nakamura. Exact cylindrical soliton solutions of the sine - Gordon
	equation, the sinh - Gordon equation and the periodic Toda equation.
	J. Phys. Society Japan 57 (1988) 3309 - 3322.
\bibitem{vit94b}
	N. K. Martinov, N. K.  Vitanov. On the self - consistent thermal equilibrium
	structures in two - dimensional negative temperature systems. Canadian
	Journal of Physics 72 (1994) 618 - 624.
\bibitem{wazwaz06}
	A. M. Wazwaz. Exact solutions for the generalized sine - Gordon and 
	sinh-Gordon equation. Chaos Solitons \& Fractals 28 (2006) 127 - 135.
\bibitem{vit07}
	S. Panchev, T. Spassova, N. K.  Vitanov. Analytical and numerical
	investigation of two families of Lorenz-like dynamical systems.
	Chaos Solitons \& Fractals 33 (2007) 1658 - 1671.
\bibitem{kudr08}
	N. A. Kudryashov, N. B. Loguinova. Extended simplest equation method
	for nonlinear differential equations. Applied Mathematics and
	Computation  205 (2008) 396 - 402.
\bibitem{kudr07x}
	N. A. Kudryashov, M. V. Demina. Polygons of differential equations for
	finding exact solutions. Chaos Solitons \& Fractals 33 (2007) 480 - 496.
\bibitem{vit10_1}
	N. K. Vitanov. Application of simplest equations of Bernoulli and Riccati kind for
	obtaining exact traveling-wave solutions for a class of PDEs with polynomial
	nonlinearity. Communications in Nonlinear Science and Numerical Simulation
	15 (2010) 2050 - 2060.
\bibitem{vit11a} 
	N. K. Vitanov, Z. I. Dimitrova, K. N.  Vitanov. On the class of nonlinear
	PDEs that can be treated by the modified method of simplest equation. Application to
	generalized  Degasperis - Processi equation and $b$-equation. Communications in
	Nonliner Science and Numerical Simulation (in press), doi:10.1016/j.cnsns.2010.11.013  
\bibitem{benney}
	D. J. Benney. Long nonlinear waves in fluid flows. Journal of
	Mathematics and Physics 45 (1966) 52 - 63.
\bibitem{stocker}
	J. J. Stoker. Water waves. the mathematical theory with applications. Wiley, New York, 
	1992.
\bibitem{j1}
	R. S. Johnson. A modern introduction to the mathematical theory of water waves. Cambridge
	University Press, Cambridge, UK, 1997.
\bibitem{j2}
	R. S. Johnson. The classical problem of water waves: a reservoir of integrable and nearly 
	integrable equations. Journal of Nonlinear Mathematical Physics 10 (2003) Supplement 1, 
	72 - 92.
\bibitem{ch1}
    R. Camassa, D. D. Holm. An integrable shallow water equation with peaked solitons. Phys. 
	Rev. Lett. 71 (1993) 1661 - 1664.
\bibitem{tw}
    L. Tian, Y. Wang. Global conservative solutions of the generalized Camassa - Holm 
	equation. International Journal of Nonlinear Science 5 (2008)  195 - 202.
\bibitem{ts}
    L. Tian, X. Song. New peaked solitary wave solutions of the generalized Camassa-Holm 
	equation. Chaos Solitons \& Fractals  19 (2004) 621 - 637.
\bibitem{ty}
    L. Tian, J. Yin. New compacton solutions and solitary wave  solutions of fully nonlinear 
	generalized Camassa-Holm equations. Chaos Solitons \& Fractals 20 (2004) 289 - 299.
\bibitem{k1}
	N. A. Kudryashov, N. B. Loguinova. Be careful with the exp - function
	method. Commun. Nonlinear Sci. Numer. Simulat. 14 (2009) 1881 - 1890.
\bibitem{k2}
	N. A. Kudryashov. Seven common errors in finding exact solutions of
	nonlinear differential equations. Commun. Nonlinear
	Sci. Numer. Simulat. 14 (2009) 3507 - 3529.
\end{thebibliography}
\end{document}